\documentstyle[preprint,aps]{revtex}
\preprint{USM-TH-152}
\begin{document}
\title{Confinement in the presence of external fields and axions}
\author{P. Gaete$^{1}$\thanks{
E-mail: patricio.gaete@fis.utfsm.cl} and E. I. Guendelman
$^{2}$\thanks{ E-mail: guendel@bgumail.bgu.ac.il}}
\address{$^1$Departamento de F\'{\i}sica, Universidad T\'{e}cnica
F. Santa Mar\'{\i}a, Casilla 110-V, Valpara\'{\i}so, Chile \\
$^2$Physics Department, Ben Gurion University, Beer Sheva 84105,
Israel} \maketitle
\begin{abstract}
For a theory with a pseudo scalar coupling $\phi F\tilde F$ and in
the case that there is a constant electric or magnetic strength
expectation value, we compute the interaction potential within the
structure of the gauge-invariant but path-dependent variables
formalism. While in the case of a constant electric field strength
expectation value the static potential remains Coulombic, in the
case of a constant magnetic field strength the potential energy is
the sum of a Yukawa and a linear potentials, leading to the
confinement of static charges.
\end{abstract}
\smallskip

PACS number(s): 11.10.Ef, 11.15.-q

\section{Introduction}

One of the key issues facing $QCD$ is understanding confinement of
quarks and gluons. In fact, a linearly increasing quark-antiquark
pair static potential provides the simplest criterion for
confinement, although unfortunately there is up to now no known
way to analytically derive the confining potential from first
principles. However, as is well known, it has been approached from
many different techniques and ideas, like lattice gauge
theories\cite{Wilson}, non perturbative solutions of
Schwinger-Dyson's equations\cite{Zachariasen}. Other authors also
associate confinement with the existence of a non trivial vacuum
structure, where the chromomagnetic field strength acquires a non
zero expectation value\cite{Matinian}.

In fact non vanishing expectation values can have dramatic
consequences in everything that concerns the infrared properties
of a theory. An interesting model where these effects have been
studied to a certain extent is the "axion-gauge field" system,
where a scalar field $\phi$ (the "axion") is coupled to gauge
fields via the interaction term
\begin{equation}
{\cal L}_I  = \frac{g}{8}\phi \varepsilon ^{\mu \nu \alpha \beta }
F_{\mu \nu } F_{\alpha \beta }. \label{GM1}
\end{equation}
This theory experiences mass generation when $\phi$ develops a
space dependent expectation value\cite{Owen} and tachyonic mass
generation when a time dependent expectation value
appears\cite{Field}. Mass generation is also achieved when the
gauge field $F_{\mu \nu }$ takes a magnetic type expectation
value\cite{Spallucci}. If $F_{\mu \nu }$ takes an electric type
expectation value, tachyonic mass generation takes
place\cite{Spallucci}. Thus, in order to gain further insight into
the physics presented by this theory, in this paper we will focus
attention on the static potential between charged fields. The
purpose here is to investigate the effects of the external
expectation value field strength on the interaction energy.

The interaction energy between static charges is a tool of
considerable interest which is expected to provide the foundation
for understanding confinement, and its physical content can be
understood when a correct separation of the physical degrees of
freedom is made. Previously, we proposed a general framework for
studying the confining and screening nature of the static
potential in gauge theories in terms of the gauge-invariant but
path-dependent field variables\cite{Pato}. An important feature of
this methodology is that it provides a physically-based
alternative to the usual Wilson loop approximation. When we
compute in this way the static potential for the model described
in \cite{Spallucci}, which contains the term (\ref{GM1}), in the
presence of an external field strength which can be either
electric or magnetic, the result of this calculation is rather
unexpected in the magnetic case: It is shown that the interaction
energy is the superposition of a Yukawa and a linear potentials,
that is, the confinement between static charges is obtained. On
the other hand, in the case of a constant electric field strength
expectation value the static potential remains Coulombic, that is,
the interaction energy does not exhibit any sensitive
modification. Actually, the linear confining potential seems to be
associated only with the magnetic field strength expectation
value.

\section{Interaction energy}

Before going to the derivation of the interaction energy, we will
describe very briefly the model under consideration. We start from
the following effective Lagrangian \cite{Spallucci}:
\begin{equation}
{\cal L} = - \frac{1}{4}f_{\mu \nu } f^{\mu \nu }  - \frac{{g^2
}}{{16}}\varepsilon ^{\mu \nu \alpha \beta } \left\langle {F_{\mu
\nu } } \right\rangle \varepsilon ^{\rho \sigma \gamma \delta }
\left\langle {F_{\rho \sigma } } \right\rangle f_{\alpha \beta }
\frac{1}{{\Box  + m_A^2 }}f_{\gamma \delta }.  \label{GM2}
\end{equation}
where $\left\langle {F_{\mu \nu } } \right\rangle$ represents the
constant classical background (which is a solution of the
classical equations of motion), and $m_A$ is the mass for the
axion field. Here, $f_{\mu \nu }  =\partial _\mu A_\nu -\partial
_\nu A_\mu$ describes a small fluctuation around the background.
We also mention that the above Lagrangian arose after using $
\varepsilon ^{\mu \nu \alpha \beta } \left\langle {F_{\mu \nu } }
\right\rangle \left\langle {F_{\alpha \beta } } \right\rangle=0$
(which holds for a pure electric or a pure magnetic background),
and integrating  out the axion fields $\phi$.

By introducing $\varepsilon ^{\mu \nu \alpha \beta } \left\langle
{F_{\mu \nu } } \right\rangle  \equiv v^{\alpha \beta }$ and $
\varepsilon ^{\rho \sigma \gamma \delta } \left\langle {F_{\rho
\sigma } } \right\rangle  \equiv v^{\gamma \delta }$, it follows
that the expression (\ref{GM2}) can be rewritten as
\begin{equation}
{\cal L} =  - \frac{1}{4}f_{\mu \nu } f^{\mu \nu }  - \frac{{g^2
}}{{16}}v^{\alpha \beta } f_{\alpha \beta } \frac{1}{{ \Box  +
m_A^2 }}v^{\gamma \delta } f_{\gamma \delta }, \label{GM3}
\end{equation}
still, the tensor $v^{\alpha \beta }$ is not arbitrary, but must
satisfy $\varepsilon ^{\mu \nu \alpha \beta } v_{\mu \nu }
v_{\alpha \beta }=0$.

\subsection{Magnetic case}

As stated, our main objective is to calculate the interaction
energy in the $v^{0i} \ne 0$ and $v^{ij}=0$ case (referred to as
the magnetic one in what follows), following the conventional path
via the expectation value of the Hamiltonian in the physical state
$\left| \Phi  \right\rangle$, which we will denote by $
\left\langle H \right\rangle _\Phi$. The Lagrangian (\ref{GM3})
then becomes
\begin{equation}
{\cal L} =  - \frac{1}{4}f_{\mu \nu } f^{\mu \nu }  - \frac{{g^2
}}{{16}}v^{0i} f_{0i} \frac{1}{{ \Box  + m_A^2 }}v^{0k} f_{0k} -
A_0 J^0,  \label{GM4}
\end{equation}
where $J^0$ is an external current, $(\mu ,\nu  = 0,1,2,3)$ and
$(i,k= 1,2,3)$.

We now proceed to obtain the Hamiltonian. For this we consider the
Hamiltonian formulation of this theory. The canonical momenta
obtained from (\ref{GM4}) are
\begin{equation}
\Pi^0=0, \label{GM41}
\end{equation}
and
\begin{equation}
\Pi _i  = D_{ij} E_j, \label{GM42}
\end{equation}
where $E_i  \equiv F_{i0}$ and $D_{ij}  \equiv \left( {\delta
_{ij}  - \frac{{g^2 }}{8}v_{i0} \frac{1}{{ \Box  + m_A^2 }}v_{j0}
} \right)$. Since $D$ is a nonsingular matrix $(\det D = 1 -
\frac{{g^2 }}{8}\frac{{{\bf v}^2 }}{{ \Box  + m_A^2 }} \ne 0)$
with ${\bf v}^2  \equiv v^{i0} v^{i0}$, there exists the inverse
of $D$ and from Eq.(\ref{GM42}) we obtain
\begin{equation}
E_i  = \frac{1}{{\det D}}\left\{ {\delta _{ij} \det D + \frac{{g^2
}}{8}v_i \frac{1}{{ \Box  + m_A^2 }}v_j } \right\}\Pi _j.
\label{GM5}
\end{equation}

The canonical Hamiltonian corresponding to (\ref{GM4}) is
\begin{equation}
H_C  = \int {d^3 } x\left\{ { - A_0 \left( {\partial _i \Pi ^i  -
J^0 } \right) + \frac{1}{2}{\bf \Pi} ^2  + \frac{{g^2
}}{{16}}\frac{{\left( {{\bf v} \cdot {\bf \Pi} } \right)^2
}}{{\left( { \Box  + M^2 } \right)}} + \frac{1}{2}{\bf B}^2 }
\right\}, \label{GM6}
\end{equation}
where $M^2  \equiv m_A^2  - \frac{{g^2 }}{8}{\bf v}^2$ and ${\bf
B}$ is the magnetic field. Demanding that the primary constraint $
\Pi _0=0$ be preserved in the course of time, one obtains the
secondary Gauss law constraint of the theory as $\Gamma _1 \left(
x \right) \equiv \partial _i \Pi ^i - J^0 = 0$. The preservation
of $\Gamma_1$ for all times does not give rise to any further
constraints. The theory is thus seen to possess only two
constraints, which are first class, therefore the theory described
by $(\ref{GM4})$ is a gauge-invariant one. The extended
Hamiltonian that generates translations in time then reads $H =
H_C  + \int {d^3 } x\left( {c_0 \left( x \right)\Pi _0 \left( x
\right) + c_1 \left( x \right)\Gamma _1 \left( x \right)}
\right)$, where $c_0 \left( x \right)$ and $c_1 \left( x \right)$
are the Lagrange multiplier fields. Moreover, it is
straightforward to see that $\dot{A}_0 \left( x \right)= \left[
{A_0 \left( x \right),H} \right] = c_0 \left( x \right)$, which is
an arbitrary function. Since $ \Pi^0 = 0$ always, neither $ A^0 $
nor $ \Pi^0 $ are of interest in describing the system and may be
discarded from the theory. Thus the Hamiltonian takes the form
\begin{equation}
H = \int {d^3 x} \left\{ {\frac{1}{2}{\bf \Pi} ^2  + \frac{{g^2
}}{{16}}\frac{{\left( {{\bf v} \cdot {\bf \Pi} } \right)^2 }}{{
(\Box + M^2) }} + \frac{1}{2}{\bf B}^2  + c(x)\left( {\partial _i
\Pi ^i - J^0 } \right)} \right\}, \label{GM7}
\end{equation}
where $c(x) = c_1 (x) - A_0 (x)$.

To quantize the theory using Dirac's procedure\cite{Dirac} we
introduce a supplementary condition on the vector potential such
that the full set of constraints becomes second class. For this
purpose, we could choose, for example, the gauge-fixing condition
\cite{Pato}
\begin{equation}
\Gamma _2 \left( x \right) \equiv \int\limits_{C_{\xi x} } {dz^\nu
} A_\nu \left( z \right) \equiv \int\limits_0^1 {d\lambda x^i }
A_i \left( {\lambda x} \right) = 0, \label{GM7}
\end{equation}
where  $\lambda$ $(0\leq \lambda\leq1)$ is the parameter
describing the spacelike straight path $ x^i = \xi ^i  + \lambda
\left( {x - \xi } \right)^i $, and $ \xi $ is a fixed point
(reference point). There is no essential loss of generality if we
restrict our considerations to $ \xi ^i=0 $. In this case, the
only nonvanishing equal-time Dirac bracket is
\begin{equation}
\left\{ {A_i \left( x \right),\Pi ^j \left( y \right)} \right\}^ *
=\delta{ _i^j} \delta ^{\left( 3 \right)} \left( {x - y} \right) -
\partial _i^x \int\limits_0^1 {d\lambda x^j } \delta ^{\left( 3
\right)} \left( {\lambda x - y} \right). \label{GM8}
\end{equation}
In passing we recall that the transition to quantum theory is made
by the replacement of the Dirac brackets by the operator
commutation relations according to
\begin{equation}
\left\{ {A,B} \right\}^ *   \to \left( { - i} \right)\left[ {A,B}
\right]. \label{Brac}
\end{equation}

We are now in a position to evaluate the interaction energy
between pointlike sources in the model under consideration, where
a fermion is localized at ${\bf y}\prime$ and an antifermion at $
{\bf y}$. From our above discussion, we see that $\left\langle H
\right\rangle _\Phi$ reads
\begin{equation}
\left\langle H \right\rangle _\Phi   = \left\langle \Phi
\right|\int {d^3 x} \left\{ {\frac{1}{2}{\bf \Pi} ^2  + \frac{{g^2
}}{{16}}\frac{{\left( {{\bf v} \cdot {\bf \Pi} } \right)^2 }}{{
(\Box + M^2) }} + \frac{1}{2}{\bf B}^2 } \right\}\left| \Phi
\right\rangle. \label{GM9}
\end{equation}
Next, as was first established by Dirac\cite{Dirac2}, the physical
state can be written as
\begin{equation}
\left| \Phi  \right\rangle  \equiv \left| {\overline \Psi  \left(
\bf y \right)\Psi \left( {\bf y}\prime \right)} \right\rangle  =
\overline \psi \left( \bf y \right)\exp \left(
{ie\int\limits_{{\bf y}\prime}^{\bf y} {dz^i } A_i \left( z
\right)} \right)\psi \left({\bf y}\prime \right)\left| 0
\right\rangle, \label{GM10}
\end{equation}
where $\left| 0 \right\rangle$ is the physical vacuum state and
the line integral appearing in the above expression is along a
spacelike path starting at ${\bf y}\prime$ and ending at $\bf y$,
on a fixed time slice. From this we see that the fermion fields
are now dressed by a cloud of gauge fields. As mentioned before,
the fermions are taken to be infinitely massive (static).
Consequently, we can write Eq.(\ref{GM9}) as
\begin{equation}
\left\langle H \right\rangle _\Phi   = \left\langle \Phi
\right|\int {d^3 x} \left\{ {\frac{1}{2}{\bf \Pi} ^2  - \frac{{g^2
}}{{16}}\frac{{\left( {{\bf v} \cdot {\bf \Pi} } \right)^2
}}{{\nabla ^2 - M^2 }}} \right\}\left| \Phi  \right\rangle,
\label{GM11}
\end{equation}
with $\partial _i \partial ^i  =  - \nabla ^2$.

From the foregoing Hamiltonian discussion, we first note that
\begin{equation}
\Pi _i(\bf x) \left| {\overline \Psi  \left( {\bf y} \right)\Psi
\left( {{\bf y}^\prime } \right)} \right\rangle  = \overline \Psi
\left( {\bf y} \right)\Psi \left( {{\bf y}^\prime  } \right) \Pi
_i(\bf x)\left| 0 \right\rangle + e\int_{\bf y}^{{\bf y}\prime} d
z_i \delta ^{\left( 3 \right)} \left( {{\bf z} - {\bf x}}
\right)\left| \Phi \right\rangle. \label{GM12}
\end{equation}
Combining Eqs.(\ref{GM11}) and (\ref{GM12}), we have
\begin{equation}
\left\langle H \right\rangle _\Phi   = \left\langle H
\right\rangle _0  + V_1  + V_2, \label{GM13}
\end{equation}
where $\left\langle H \right\rangle _0  = \left\langle 0
\right|H\left| 0 \right\rangle$.

The $V_1$ term is given by
\begin{equation}
V_1  = \frac{{e^2 }}{2}\int_{\bf y}^{{\bf y}^{\prime}  }
{dz_i^{\prime}}\partial _i^{z^{\prime}} \int_{\bf y}^{{\bf
y}^{\prime}} {dz^i }\partial _z^i G\left( {{\bf z}^{\prime},{\bf
z}} \right), \label{GM14}
\end{equation}
where $G$ is the Green function
\begin{equation}
G({\bf z}^{\prime}  ,{\bf z}) = \frac{1}{{4\pi }}\frac{{e^{ -
M|{\bf z}^{\prime} - {\bf z}|} }}{{|{\bf z}^{\prime} - {\bf z}|}}.
\label{GM15}
\end{equation}
By means of Eq.(\ref{GM15}) and remembering that the integrals
over $z^i$ and $z_i^{\prime}$ are zero except on the contour of
integration, the term (\ref{GM14}) reduces to the Yukawa-type
potential after subtracting the self-energy terms, that is,
\begin{equation}
V_1  =  - \frac{{e^2 }}{{4\pi }}\frac{{e^{ - M|{\bf y} - {\bf y}^
{\prime}| } }}{{|{\bf y} - {\bf y}^{\prime}|}}. \label{GM16}
\end{equation}

We now come to the $V_2$ term, which is given by
\begin{equation}
V_2  = \frac{{e^2 m_A^2 }}{2}\int_{\bf y}^{{\bf y}^{\prime}  }
{dz^{{\prime} i} } \int_{\bf y}^{{\bf y}^{\prime}  } {dz^i }
G({\bf z}^{\prime} ,{\bf z}). \label{GM17}
\end{equation}
In order to compute $V_2$, we make use of the Green function
(\ref{GM15}) in momentum space
\begin{equation}
\frac{1}{{4\pi }}\frac{{e^{ - M|{\bf z}^{\prime}   - {\bf z}|}
}}{{|{\bf z}^{\prime} - {\bf z}|}} = \int {\frac{{d^3 k}}{{\left(
{2\pi } \right)^3 }}\frac{{e^{i{\bf k} \cdot \left( {{\bf
z}^{\prime}- {\bf z}} \right)} }}{{{\bf k}^2  + M^2 }}}.
\label{GM18}
\end{equation}
Thus, by employing relation (\ref{GM18}) we can reduce
Eq.(\ref{GM17}) to
\begin{equation}
V_2  = e^2 m_A^2 \int {\frac{{d^3 k}}{{\left( {2\pi } \right)^3
}}} \left[ {1 - \cos \left( {{\bf k} \cdot {\bf r}} \right)}
\right]\frac{1}{{({\bf k}^2  + M^2) }}\frac{1}{{\left( {{\bf {\hat
n}} \cdot {\bf k}} \right)^2 }}, \label{GM19}
\end{equation}
where ${\bf {\hat n}} \equiv \frac{{{\bf y} - {\bf
y}^{\prime}}}{{|{\bf y} - {\bf y}^{\prime}| }}$ is a unit vector
and ${\bf r}={\bf y}-{\bf y^{\prime}}$ is the relative vector
between the quark and antiquark. Since ${\bf {\hat n}}$ and ${\bf
r}$ are parallel, we get accordingly
\begin{equation}
V_2  = \frac{{e^2 m_A^2 }}{{8\pi ^3 }}\int\limits_{ - \infty
}^\infty  {\frac{{dk_r }}{{k_r^2 }}} \left[ {1 - \cos \left( {k_r
r} \right)} \right]\int\limits_0^\infty  {d^2 k_T \frac{1}{{(k_r^2
+ k_T^2  + M^2) }}}, \label{GM20}
\end{equation}
where $k_T$ denotes the momentum component perpendicular to ${\bf
r}$. We may further simplify Eq.(\ref{GM20}) by doing the $k_T$
integral, which leads immediately to the result
\begin{equation}
V_2  = \frac{{e^2 m_A^2 }}{{8\pi ^2 }}\int\limits_{ - \infty
}^\infty  {\frac{{dk_r }}{{k_r^2 }}} \left[ {1 - \cos \left( {k_r
r} \right)} \right]\ln \left( {1 + \frac{{\Lambda ^2 }}{{k_r^2  +
M^2 }}} \right), \label{GM21}
\end{equation}
where $\Lambda$ is an ultraviolet cutoff. We also observe at this
stage that similar integral was obtained independently in
Ref.\cite{Suganuma} in the context of the dual Ginzburg-Landau
theory by an entirely different approach.

Now, we move on to compute the integral (\ref{GM21}). To this end
it is advantageous to introduce a new auxiliary parameter
$\varepsilon$ by making in the denominator of the integral
(\ref{GM21}) the substitution $k_r^2\rightarrow
k_r^2+\varepsilon^2$. This allows us to obtain a form more
comfortable to handle the integral. Hence we evaluate $\lim
_{\varepsilon \to 0} {\widetilde V}_2$, that is,
\begin{equation}
V_2\equiv\lim _{\varepsilon  \to 0} {\widetilde V}_2= \lim
_{\varepsilon  \to 0}\frac{{e^2 m_A^2 }}{{8\pi ^2 }}\int\limits_{
- \infty }^\infty {\frac{{dk_r }}{{(k_r^2  + \varepsilon ^2) }}}
\left[ {1 - \cos \left( {k_r r} \right)} \right]\ln \left( {1 +
\frac{{\Lambda ^2 }}{{k_r^2  + M^2 }}} \right). \label{GM22}
\end{equation}
The integration on the $k_r$-complex plane yields
\begin{equation}
{\widetilde V}_2 = \frac{{e^2 m_A^2 }}{{8\pi }}\left( {\frac{{1 -
e^{ - \varepsilon |{\bf y} - {{\bf y}^\prime}|  } }}{\varepsilon
}} \right)\ln \left( {1 + \frac{{\Lambda ^2 }}{{M^2 - \varepsilon
^2 }}} \right). \label{GM23}
\end{equation}
Taking the limit $\varepsilon  \to 0$, expression (\ref{GM23})
then becomes
\begin{equation}
V_2  = \frac{{e^2 m_A^2 }}{{8\pi }}|{\bf y} - {{\bf y}^\prime}|
\ln \left( {1 + \frac{{\Lambda ^2 }}{{M^2 }}} \right).
\label{GM24}
\end{equation}
This, together with Eq.(\ref{GM16}), yields finally
\begin{equation}
V(L) =  - \frac{{e^2 }}{{4\pi }}\frac{{e^{ - ML} }}{L} +
\frac{{e^2 m_A^2 }}{{8\pi }}L\ln \left( {1 + \frac{{\Lambda ^2
}}{{M^2 }}} \right), \label{GM25}
\end{equation}
where $L\equiv|{\bf y}-{\bf {y^\prime}}|$.

It is worth noting here that this is exactly the result obtained
in Ref.\cite{Suganuma} in the context of the dual Landau-Ginzburg
thaeory. But we do not think that the agreement is an accidental
coincidence. Also, the massive Abelian antisymmetric tensor gauge
theory displays the same behavior\cite{Deguchi,Pato3}. In other
words, there is a class of models which can predict this
interaction energy.

\subsection{Electric case}

We now want to extend what we have done to the case $v^{0i}=0$ and
$v^{ij}\ne 0$ (referred to as the electric one in what follows).
In such a case the Lagrangian reads
\begin{equation}
{\cal L} =  - \frac{1}{4}f_{\mu \nu } f^{\mu \nu }  - \frac{{g^2
}}{{16}}v^{ij} f_{ij} \frac{1}{{ \Box  + m_A^2 }}v^{kl} f_{kl} -
A_0 J^0, \label{GM26}
\end{equation}
$(\mu ,\nu  = 0,1,2,3)$ and $(i,j,k,l = 1,2,3)$.

The above Lagrangian will be the starting point of the Dirac
constrained analysis. The canonical momenta following from
Eq.(\ref{GM26}) are $\Pi^\mu=f^{\mu0}$, which results in the usual
primary constraint $\Pi^0=0$ and $\Pi^i=f^{i0}$. Defining the
electric and magnetic fields by $ E^i  = F^{i0}$ and $B^i  =
\frac{1}{2}\varepsilon ^{ijk} F_{jk}$, respectively, the canonical
Hamiltonian assumes the form
\begin{equation}
H_C  = \int {d^3 } x\left\{ {\frac{1}{2}{\bf E}^2+\frac{1}{2}{\bf
B}^2+\frac{{g^2}}{{16}}\varepsilon_{ijm}\varepsilon_{kln}v^{ij}
B^{m} \frac{1}{{ \Box  + m_A^2 }}v^{kl} B^{n}  - A_0 \left(
{\partial _i \Pi ^i - J^0 } \right)} \right\}.  \label{GM27}
\end{equation}
Time conservation of the primary constraint leads to the secondary
constraint $\Gamma_1(x) \equiv \partial_i\Pi^i - J^0=0$, and the
time stability of the secondary constraint does not induce more
constraints, which are first class. It should be noted that the
constrained structure for the gauge field is identical to the
usual Maxwell theory. Notwithstanding, in order to put our
discussion into context it is useful to summarize the relevant
aspects of the analysis described previously\cite{Pato2}. In view
of this situation, we pass now to the calculation of the
interaction energy.

Following our earlier procedure, we will compute the expectation
value of the Hamiltonian in the physical state $\left| \Phi
\right\rangle$ (Eq. (\ref{GM10})). That is,
\begin{equation}
 \left\langle H \right\rangle _\Phi   = \left\langle \Phi
\right|\int {d^3 x} \left\{ {\frac{1}{2}{\bf E}^2 } \right\}\left|
\Phi \right\rangle. \label{GM28}
\end{equation}
Taking into account the above Hamiltonian structure, the
interaction takes the form
\begin{equation}
\langle H\rangle _{\Phi }=\langle H\rangle _{0}+\frac{e^{2}}{2}\int_{%
\bf{y}}^{\bf{y}^{\prime }}dz^{i}\int_{\bf{y}}^{\bf{y}%
^{\prime }}dz_{i}^{\prime }\text{ }\delta ^{(3)}(\bf{z}-\bf{z}%
^{\prime }),  \label{GM29}
\end{equation}
where $\langle H\rangle _{0}=\langle 0\mid H\mid 0\rangle$. Once
again, the integrals over $z_{i}$ and $z_{i}^{\prime }$ are zero
except on the contour of integrations, accordingly one obtains the
following interaction energy:
\begin{equation}
V=\frac{e^{2}}{2}\text{ }k\mid \bf{y}-\bf{y}^{\prime }\mid \text{
}, \label{GM30}
\end{equation}
where $k=\delta ^{(2)}(0)$. This expression shows that special
care has to be exercised in order to clarify the appearance of
this peculiar result, as was discussed elaborately in
\cite{Pato2}. It may be recalled, however, that the origin of the
divergence is quite clear, so that it is possible to extract the
Coulomb potential from the infinite contribution. Notice that the
origin of the divergent factor $k$ is due to the fact that the
thickness of the string is nonvanishing only on the contour of
integration. We recall that a suitable examination of the term
$\frac{ e^{2}}{2}\int d^{3}x\left( \int_{\bf{y}}^{\bf{y\prime
}}dz_{i}\delta ^{(3)}(\bf{x-z})\right) ^{2}$ reproduces exactly
the expected Coulomb interaction between charges after subtracting
the self-energy term\cite{Pato2}, hence Eq.(\ref{GM30}) reduces to
\begin{equation}
V(L) = - \frac{1}{{4\pi }} \frac{1}{L}, \label{GM31}
\end{equation}
where $L\equiv|{\bf y}-{\bf {y^\prime}}|$.

\section{Final Remarks}

We briefly summarize the results obtained so far. By using the
gauge-invariant but path-dependent formalism, we have studied the
static potential for the system consisting of a gauge field
interacting with a massive axion field in the case when there are
nontrivial constant expectation values for the gauge field
strength $F_{\mu \nu }$. The constant gauge field configuration is
a solution of the classical equations of motion.

While in the case when $\left\langle {F_{\mu \nu } }
\right\rangle$ is electric-like no unexpected features are found,
we find that the case when $\left\langle {F_{\mu \nu } }
\right\rangle$ is magnetic-like is totally different. In fact,
when $\left\langle {F_{\mu \nu } } \right\rangle$ is
magnetic-like, the potential between static charges displays a
Yukawa piece plus a linear confining piece. Unexpectedly, a
confining potential between static charges appears in this case.
It is interesting to note that the requirement that $\left\langle
{F_{\mu \nu } } \right\rangle$ be magnetic in order to get
confining behavior coincides with ideas concerning the nature of
the $QCD$ vacuum, where a nontrivial magnetic field strength must
be present in the vacuum\cite{Matinian}.

In the non-Abelian generalization of this model similar effects
should appear.

Further investigations of the relations between our work and the
magnetic models of the $QCD$ vacuum in Ref.\cite{Matinian} have to
be performed. Also one should address the question of what is the
physical origin of the axion field, it is probably some kind of a
bound state, if our theory represents an effective approach to
$QCD$.

\section{ACKNOWLEDGMENTS}

One of us (E.I.G.) wants to thank the Physics Department of the
Universidad T\'{e}cnica F. Santa Mar\'{\i}a for hospitality. P.G.
would like to thank G. Cvetic for useful comments on the
manuscript.


\begin{thebibliography}{}
\bibitem{Wilson}  K. G. Wilson, Phys. Rev. {\bf D10}, 2445
(1974).

\bibitem{Zachariasen} M. Baker, J. S. Ball and F. Zachariasen,
Phys. Rept. {\bf 209}, 73 (1991).

\bibitem{Matinian} S. G. Matinian and G. K. Savvidy,
Nucl. Phys. {\bf B134}, 539 (1978); G. K. Savvidy, Phys. Lett.
{\bf B71}, 133 (1977); P. Olesen, Physica Scripta, {\bf 23}, 1000
(1981); H. B. Nielsen and P. Olesen, Nucl. Phys. {\bf B160}, 380
(1979); H. B. Nielsen and M. Ninomiya, Nucl. Phys. {\bf B156}, 1
(1979).

\bibitem{Owen} E. I. Guendelman and D. A. Owen, Phys. Lett. {\bf
B251}, 439 (1990); Phys. Lett. {\bf B252}, 113 (1990).

\bibitem{Field} E. I. Guendelman and D. A. Owen, Phys. Lett. {\bf
B276}, 108 (1992); R. Brustein and D. H. Oakmin, Phys. Rev. Lett.
{\bf 82}, 2628 (1999); W. D. Garretson, G. B. Field and S. M.
Carroll, Phys. Rev. {\bf D46}, 5346 (1992); G. B. Field and S. M.
Carroll, Phys. Rev. {\bf D62}, 103008 (2000); M. S. Turner and L.
M. Widrow,  Phys. Rev. {\bf D37}, 2743 (1988); S. M. Carroll, G.
B. Field and R. Jackiw,  Phys. Rev. {\bf D41}, 1231 (1990).

\bibitem{Spallucci} E. I. Guendelman, S. Ansoldi and E. Spallucci,
JHEP09, 044 (2003).

\bibitem{Pato} P. Gaete, Z. Phys. {\bf C76}, 355 (1997); Phys. Lett.
{\bf B515}, 382 (2001); Phys. Lett. {\bf B582}, 270 (2004).

\bibitem{Dirac} P. A. M. Dirac, {\it Lectures on Quantum Mechanics}
(Yeshiva University, New York, 1964).

\bibitem{Dirac2} P. A. M. Dirac, {\it The Principles of Quantum Mechanics}
(Oxford University Press, Oxford, 1958); Can. J. Phys. {\bf 33},
650 (1955).

\bibitem{Suganuma} H. Suganuma, S. Sasaki and H. Toki,  Nucl. Phys. {\bf
B435}, 207 (1995).

\bibitem{Deguchi} S. Deguchi and Y. Kokubo,  Mod. Phys. Lett. {\bf
A17}, 503 (2002).

\bibitem{Pato2} P. Gaete, Phys. Rev. {\bf D59}, 127702 (1999).

\bibitem{Pato3} P. Gaete and C. Wotzasek, "Interaction Energy
Calculation in Abelian Antisymmetric Tensor Theories" (in
preparation).
\end{thebibliography}
\end{document}